\documentclass[conference]{IEEEtran}
\IEEEoverridecommandlockouts
\usepackage{cite}
\usepackage{amsmath,amssymb,amsfonts}
\usepackage{algorithmic}
\usepackage{graphicx}
\usepackage{textcomp}
\usepackage{xcolor}
\usepackage{soul}
\newcommand{\magvector}[1]{\mathbf{#1}}

\usepackage{fancyhdr}   

\fancypagestyle{firstpage}{     
  \fancyhf{}
  \chead{\textcolor{gray}{This article has been accepted for publication in the proceedings of the \\ 7th International Conference on Activity and Behavior Computing}}
  \fancyfoot[C]{\small{\textcolor{gray}{~\copyright~ 2025 IEEE.  Personal use of this material is permitted.  Permission from IEEE must be obtained for all other uses, in any current or future media, including reprinting/republishing this material for advertising or promotional purposes, creating new collective works, for resale or redistribution to servers or lists, or reuse of any copyrighted component of this work in other works.}}}

}

\def\BibTeX{{\rm B\kern-.05em{\sc i\kern-.025em b}\kern-.08em
    T\kern-.1667em\lower.7ex\hbox{E}\kern-.125emX}}
\begin{document}

\title{Optimization of An Induced Magnetic Field-Based Positioning System\\
}

\author{
\IEEEauthorblockN{Sizhen Bian, Gerald Pirkl, Jingyuan Cheng, Paul Lukowicz}
\IEEEauthorblockA{\textit{German Research Center For Artificial Intelligence(DFKI)} \\
Kaiserslautern, Germany \\
name.surname@dfki.de}
}

\maketitle

\begin{abstract}
Using oscillating magnetic fields for indoor positioning is a robust way to resist dynamic environments. This work presents the hard- and software-related optimizations of an induced magnetic field positioning system. We describe a new coil architecture for both the transmitter and receiver, reducing inter-axes cross-talk. A new analog circuit design on the receiver side attains an acceptable noise level and increases the detection range from 4m to 8m (the covered area is increased from $50m^2$ to $200m^2$). The median positioning error is reduced from 0.56~m to 0.25m in the near field with fingerprinting methods. Experiments in office and factory areas (including robotic and industrial equipment) demonstrate the system's robustness in large areas. This work aims to enlighten researchers working on the same topic with constructive optimization directions on their own induced magnetic field-based systems.
\end{abstract}


\section{Introduction}
\thispagestyle{firstpage}

In the past decades, several indoor positioning technologies have been proposed and experimented 
with by both academia and the industry\cite{garcia2024relabeling, subedi2020survey, shang2022overview, bai2020low, garcia2024relabeling_a, bian2024body}. Although there are highly accurate positioning technologies for 
special applications, these systems are expensive and require complex deployment and calibration \cite{li2020camera, zhang2021rgb}. 
Due to the dynamic nature of indoor environments in combination with their complex layout, many 
physical modalities suffer from signal scattering, occlusion, diffraction, or reflections caused by the interaction of the signal with the environment \cite{subedi2020survey}. Those systems (for example, ultra sound-based systems \cite{kardzhiev2022new}) 
therefore perform best if direct line-of-sight between the emitters and the receivers is available. 
Thus the complexity and costs rise as this can only be 
achieved by additional hardware installations. One way to overcome the requirement of direct 
line-of-sight is the usage of signals with low environmental influence, which permeate most materials, for example, oscillating magnetic fields\cite{golestani2021wearable, bian2021induced,golestani2020human, bian2020wearable}. 
The advantage of magnetic fields as a source of information for indoor positioning is that the fields permeate non-ferromagnetic materials and low-conductive material (including human beings) without being greatly changed (in contrast to, for example, WIFI-based localization systems). 
Although ferromagnetic materials have local influences on the magnetic fields,  the 
overall structure of the fields is hardly changed. The deformations are often repeatable and can usually be modeled to reduce their effects.
Magnetic fields as underlying physical modality have already been used in literature \cite{Abrudan15,bian20233d, Pirkl13}. Most of such systems rely on the magneto-inductance of magnetic fields in the kHz frequency band 
on a receiver coil. Building upon Pirkl et al.'s work~\cite{Pirkl13}, who made the circuits publicly 
available, we worked on several hardware optimizations of both the transmitter coils and the receiver sensing and 
processing chain. The optimizations consider the reduction of receiver and transmitter side coil crosstalk and include a new filter and amplification architecture on the receiver side, which increases the sensing range from 4~m to 8m. We evaluate the optimizations in different office and industrial environments and show that local environmental magnetic field deformations can be reduced by a fingerprinting method.

Our highlights in this work are summarized as follows:
\begin{itemize}
\item We developed a new receiver-side analog circuit, which combines both the increase in sensitivity and the reduction of the noise to increase the range of a single transmitter coil. 
\item We present new transmitter and receiver coil designs lowering the cross talk of the coil's axes. This also reduces the complexity of the transmitter coil calibrations to transform the measured magnetic field-induced voltage into the distance.
\item Experiments with a single transmitter at different points in a public room are conducted to demonstrate the performance of the newly designed hardware and transceiver coils with regard to its distance estimation accuracy.
\item Evaluation of the localization system in three different environments: a social area with kitchen and furniture, a robotic lab (with humans and robots), and an industrial production line. 
\end{itemize}
\noindent
The set of experiments showed that the optimized hardware design has better performance in the detection range and the detection accuracy, namely an increase in the covered area from $50m^2$ (4m radius) to $200m^2$ (8m radius) and a lower standard deviation of 0.25m (initial design: 0.56m) distance error in the near field. The environmental effects on the magnetic field are reduced by fingerprinting algorithms, and the position error from 0.7m to 0.27m in industrial environments can be reduced.

\begin{figure*}[hbt]
\graphicspath{{./Figures/}}
\centering
\includegraphics[width=1.0\textwidth]{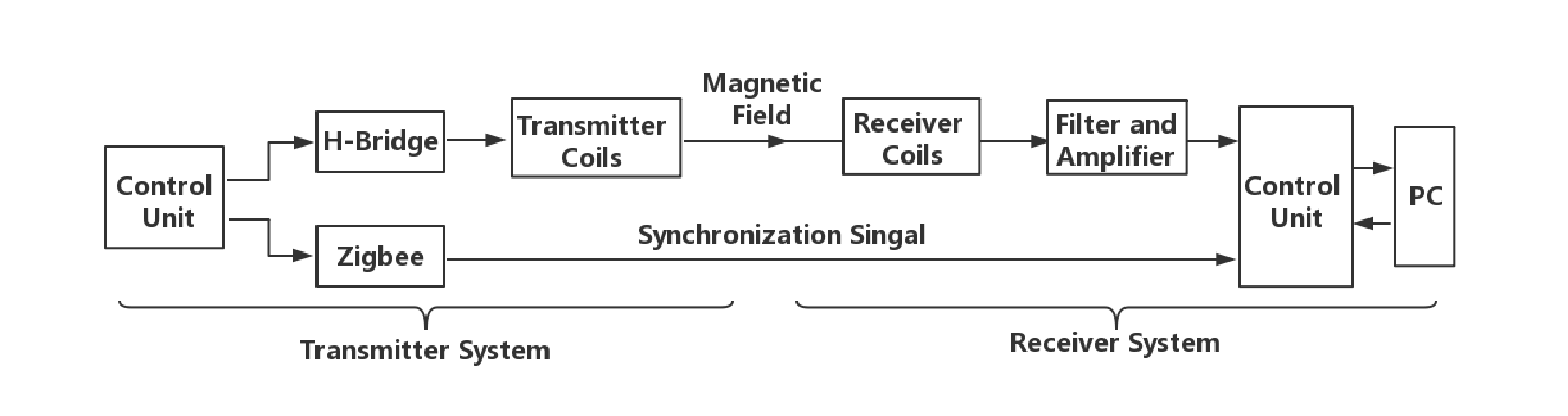}
\caption{System Architecture}
\label{fig:figure1}
\end{figure*}

\begin{figure*}[t]
\includegraphics[width=\textwidth]{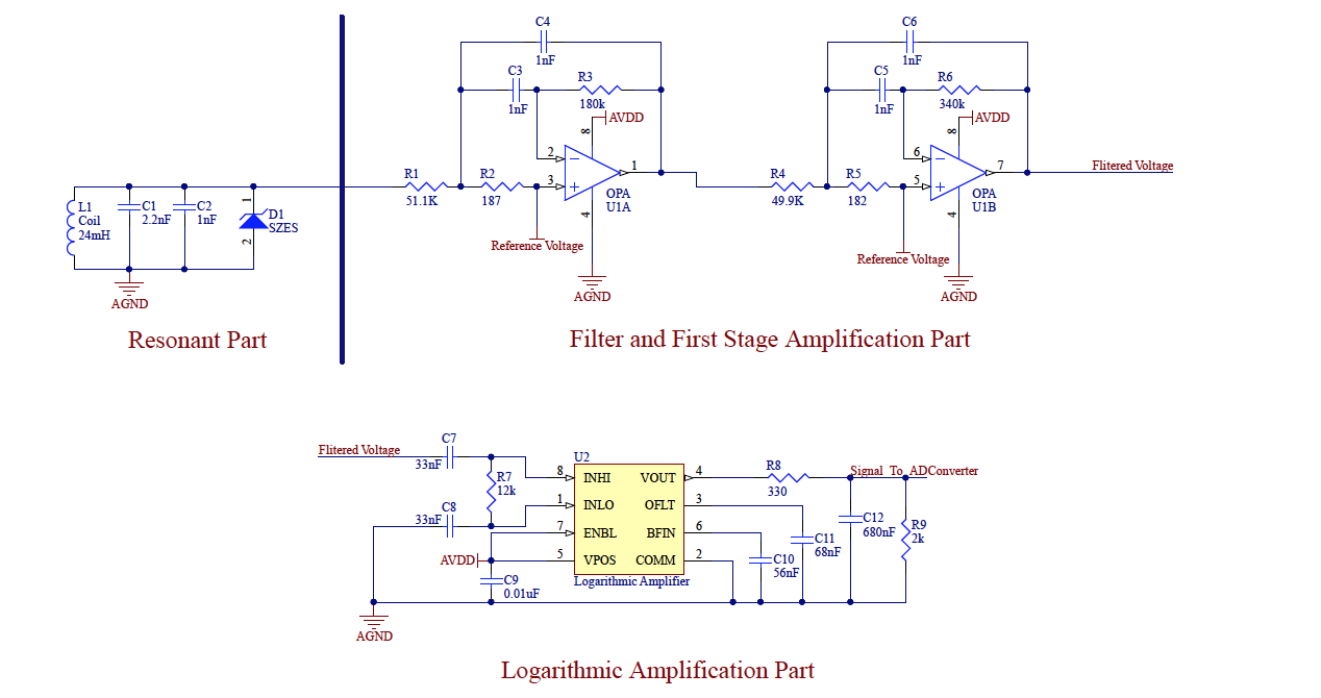}
\caption{Schematic of the Butterworth filter and logarithmic amplification chain.}
\label{fig: filter}
\end{figure*}

\section{Related Work}
This section will first give a compact summary of the latest indoor positioning modalities and then address current work on localization and position estimation with magnetic sensing technique. An extensive indoor positioning-related overview is given by Wahab~\cite{wahab2022indoor}, which is devoted to a detailed presentation of different technologies implementing indoor positioning technology. Hightower~\cite{Hightower01} presented a classification of positioning systems (triangulation, proximity, and scene analysis). They additionally present different techniques and discuss the advantages and disadvantages of the systems. Currently, available systems rely, for example, on ultra-wideband RF signals, ultrasound pulses in combination with radio frequency (RF) synchronization and data transfer~\cite{Medina13}, pedestrian dead reckoning~\cite{Woyano16}, RF fingerprinting~\cite{Beder12} or cameras~\cite{Braun13}. Also, RFID-based technologies are researched about indoor positioning, and an overview is given by Bai~\cite{Bai12}. Several competitions try to evaluate the accuracy of approaches in real-life scenarios and problems in evaluations, and the results of the competition are, for example, given in~\cite{Lymberopoulos15}. However, those implementation methods are not robust enough for a dynamic and structure-complex indoor environment. Magnetic field-based systems, especially with low frequencies, are less error-prone to multi-path phenomena and line-of-sight restrictions, which is the key advantage of a magnetic-based location system. In addition to the systems of \cite{Abrudan15} and \cite{Pirkl13}, magnetic fields are, for example, used by~\cite{Whitmire:2016:ESC:2971763.2971771} to track the eyeball of a person for augmented reality applications and by \cite{bian2020social} to measure social distancing for the interception of virus propagation. The authors of~\cite{Groben:2015:NVS:2644304.2754416} also use static magnetic fields to track the temperature distribution in beer containers during the brewing process. They additionally describe a neural network-based approach to reduce magnetic field distractions. 

Existing studies have demonstrated induced magnetic field positioning architectures using both single-axis coils and three-axis orthogonal coil configurations for spatial localization\cite{wu2018new, bian2022human, sheinker2014remote,abrudan2015distortion}. However, these traditional setups often face challenges such as cross-talk interference and non-uniform field distribution, which can degrade localization accuracy.

Previous works, such as those cited in the survey\cite{pasku2017magnetic, psiuk2019inductive}, primarily utilize square orthogonal coil configurations (similar to Fig. 4A)\cite{pirkl2012robust,abrudan2015distortion}. These designs provide a fundamental framework for indoor localization but may suffer from undesired electromagnetic coupling between coil axes, leading to positioning inaccuracies. Recognizing these limitations, this work will introduce an enhanced coil and signal processing design, optimizing field uniformity and reducing cross-talk interference to improve positioning reliability.

In addition to coil structure refinements, we also investigated analog circuit optimizations to enhance signal acquisition and processing efficiency in magnetic positioning systems. Advanced filtering techniques, noise suppression mechanisms, and amplification circuits have been proposed to strengthen signal integrity and extend the effective positioning volume\cite{de2014indoor, ouyang2022survey}. While existing magnetic field positioning architectures provide the foundation for reliable localization, our work aims to offer constructive optimization strategies that researchers can integrate into their own induced magnetic field-based systems, further advancing the accuracy.

Unlike prior works that focus on establishing positioning architectures, our contribution lies in refining existing methodologies to improve system robustness, accuracy, and scalability.

\section{System Description}

\subsection{Physical Background}
Assume that the origin of the coordinate system is at the center of the coil and the $z$ axis is the normal of the origin, the magnitude of the magnetic field, which points along the $z$ direction, is given in a simplified form by


\begin{equation}
B(\mathbf{x}) = \frac{\mu_ N a^2 I}{2\sqrt{(a^2+z^2)}^3}
\label{eq: magnetic}
\end{equation}
where $B$ is in Tesla,  $\mu=4\pi * 10^{-7}$  is the magnetic field permeability in the vacuum, $N$ is the number 
of turns of the field coil, $I$ is the current in the wire, in Amperes, $a$ is the radius of 
the coil in meters and $z$ is the axial distance in meters from the center of the coil. For a more accurate approximation of the magnetic field layout and to take the inhomogeneity of the field into account, a finite element model using Maxwell's electromagnetic theory can be applied or has to be dealt with as described~\cite{Groben:2015:NVS:2644304.2754416}.

At large distances, the degradation of the magnetic field strength is inversely proportional to 
the cube of distance. According to Maxwell's electromagnetic theory, a sinusoidal current 
flowing in the transmitter coil generates a sinusoidally varying magnetic field around the 
coil. In an oscillating magnetic field, according to Faraday's law of induction, the voltage 
accumulated around a closed circuit is proportional to the time rate of change of the magnetic 
flux it encloses. 
The strength of the signal that a receiver coil induces is proportional to the angular frequency 
and amplitude of the original current, and inversely proportional to the cube of the distance 
from the transmitter coil signal source to the measured position. 
With this physical model, a series of 3-D transmitter coils are built to generate low-frequency magnetic fields.  The local magnetic field strength corresponds to the induced voltage on the three receiver side coils. A magnetic field measurement of a receiver at position $\mathbf{x}$ 
is therefore a tensor $\magvector{\mathbf{x}}^i=(\mathbf{B}_x^i(\mathbf{x}),\mathbf{B}_y^i(\mathbf{x}),\mathbf{B}_z^i(\mathbf{x}))$ 
combining the three transmitter axes measurements of the receiver to transmitter $i$.

\subsection{Hardware Architecture}
Fig~\ref{fig:figure1} illustrates the architecture of this hardware system, consisting 
of two sub-systems, stationary transmitters, and mobile/wearable receiver nodes. The magnetic 
field transmitter sequentially generates a magnetic field on three perpendicular transmitter axes with a frequency of 20kHz. In addition to the magnetic field-derived distance information, the 3-axis coil transmitter setup provides information to restrict the position of the receiver to a small set of points within the octants around the transmitters. On the receiver side, a 3-axis receiver coil measures the induced voltage, which is then filtered, amplified, and digitalized. The data is wirelessly transferred to a processing computer for position estimation. A Zigbee-based TDMA\footnote{Time Division Multiple Access, see for example~\cite{Keshav},p. 127} scheme synchronizes the transmitter and receiver components to ensure an accurate mapping of the receiver side measurements to the transmitter axes.

\begin{figure}[hbt]
\includegraphics[width=0.50\textwidth]{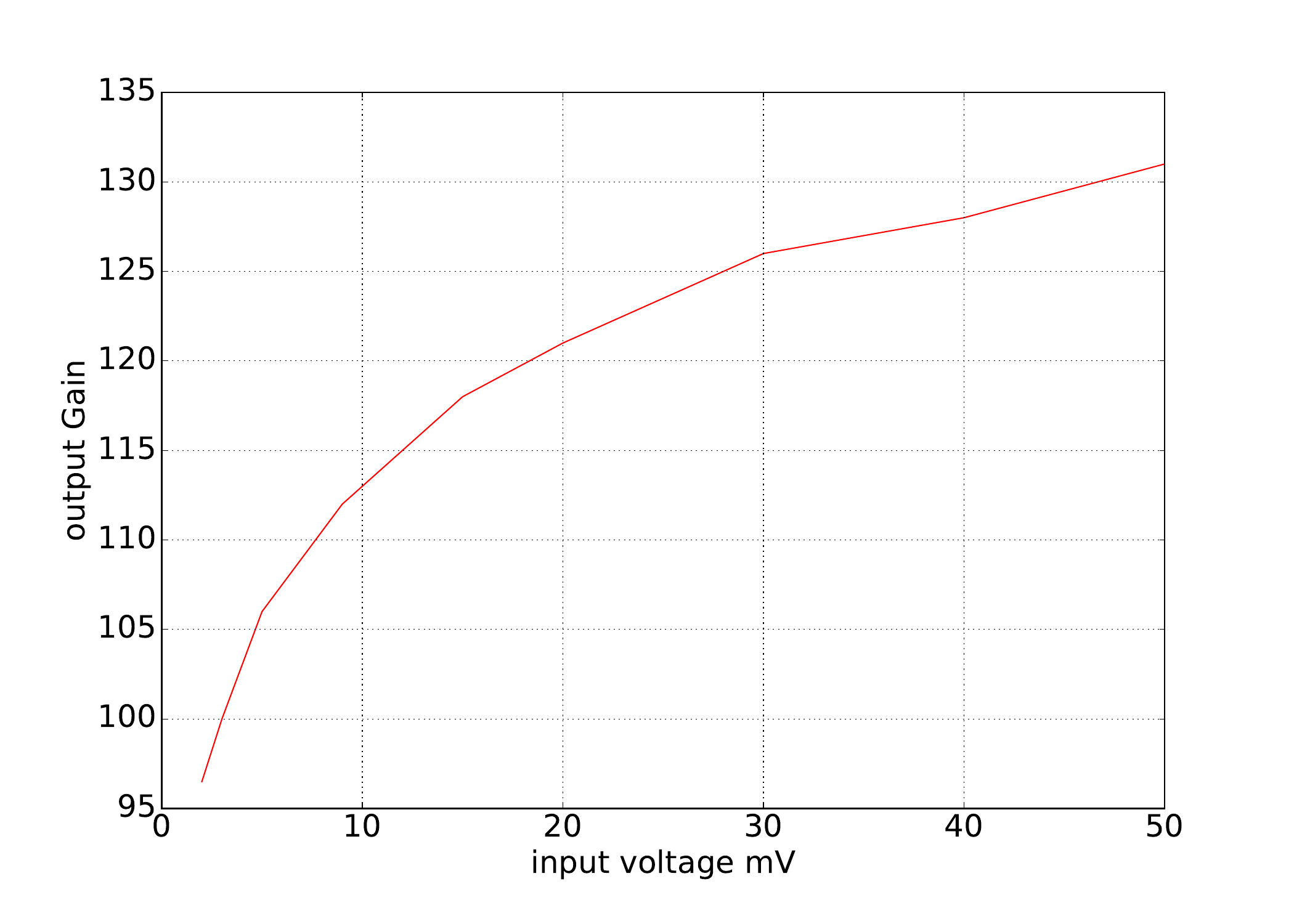}
\includegraphics[width=0.50\textwidth]{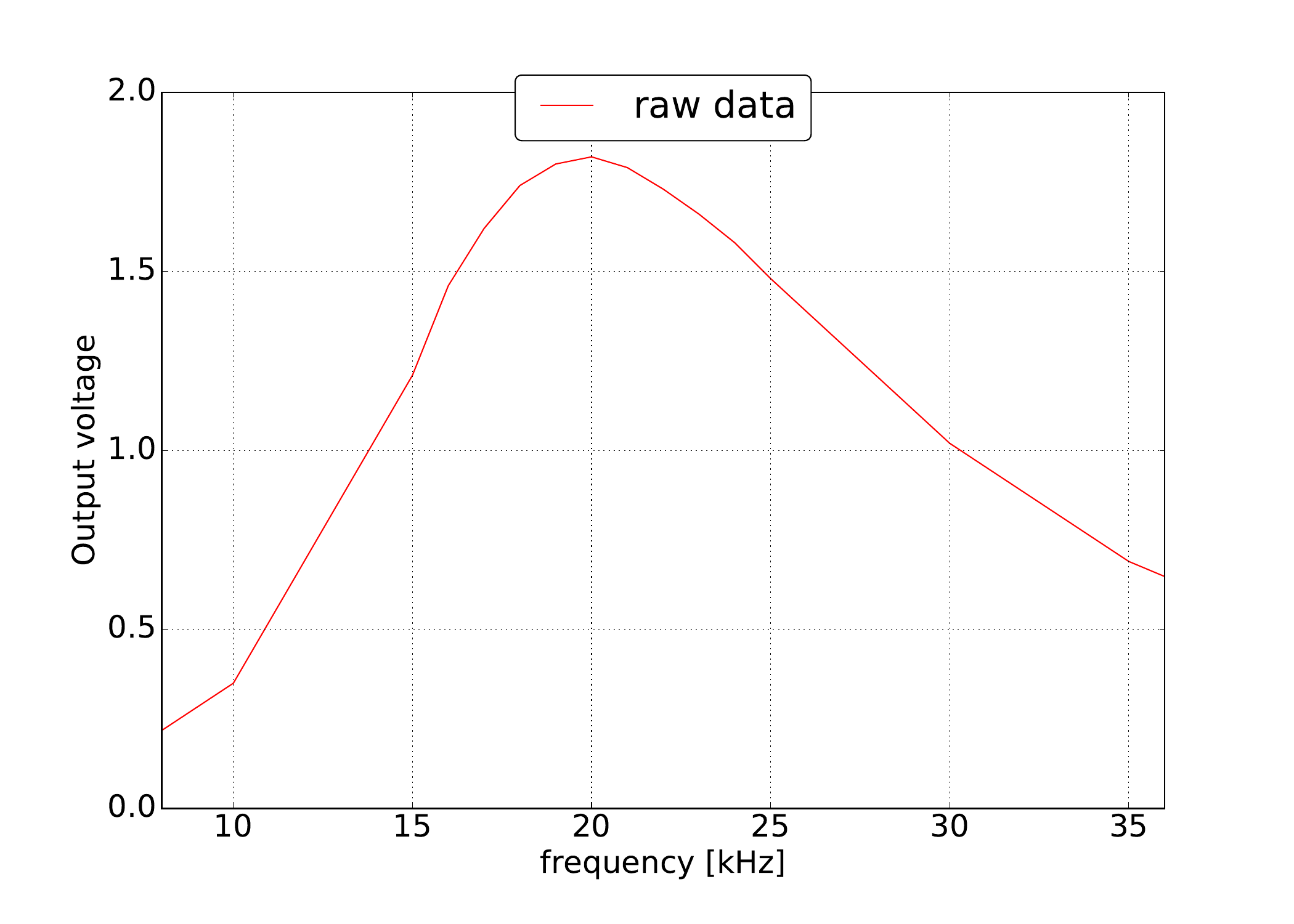}
\caption{Signal trajectories of the logarithmic amplification step and the filter response of the Butterworth filter, excitement voltage: 20mV.}
\label{fig: amp}
\end{figure}

\begin{figure*}
\centering
\includegraphics[width=0.8\textwidth]{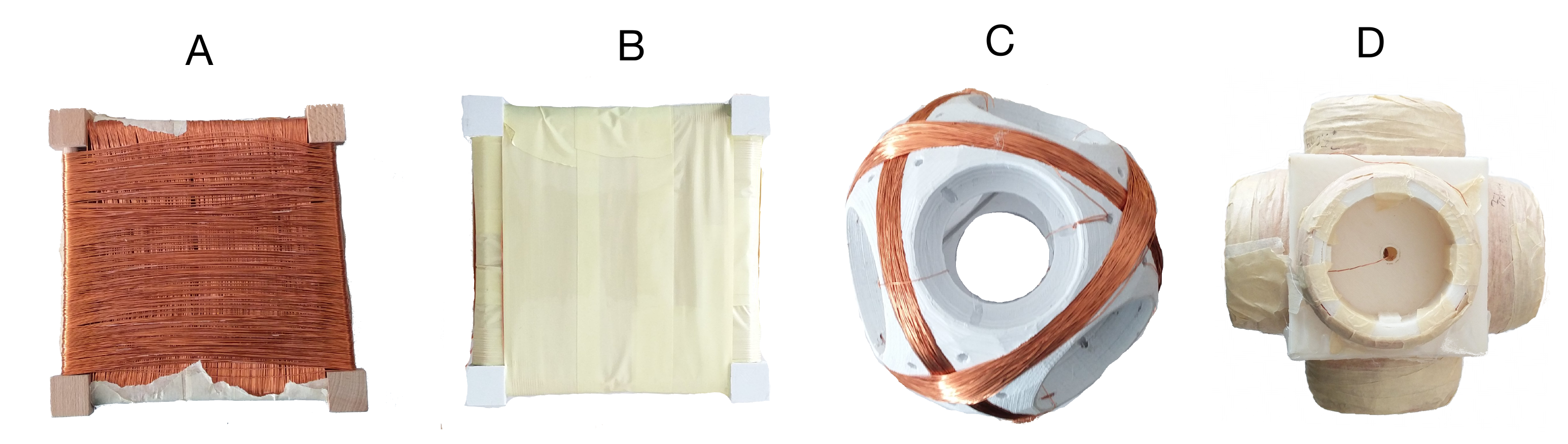}
\caption{Evaluated coils. A) and B) are the initial designs with high crosstalk due to large overlapping coil areas, C) is an intermediate version with lower cross talk due to reduced area, and D) is the final version.}
\label{fig: coils}
\end{figure*}

\subsection{Field Strength Signal Sensing}

On the receiver side,  an analog signal processing chain (see Fig. \ref{fig: filter}) filters and amplifies the voltage induced by the oscillating magnetic field.
In contrast, the board combines a filter and a logarithmic amplifier which is then followed by a high-resolution analog-to-digital converter. A 4th-order Butterworth filter is implemented to obtain amplification and bandpass filter function around the oscillating frequency of the transmitter magnetic field of 20kHz. In contrast to the previous circuit design, which consisted of a serial oscillating circuit tuned to 20kHz and an adjustable linear amplifier, the new circuit is less noisy and more accurate. 
The Butterworth response is ideal for applications requiring predictable gain characteristics, such as the anti-aliasing filter used ahead of an analog-to-digital converter. 
The filter in this system can provide a maximally flat response and high roll-off(40dB/dec). 
The main purpose of this part is to amplify the induced voltage signal and filter the signal so 
that the input signal of the logarithmic amplifier is stable and of low noise (Fig.~\ref{fig: amp} left). The essential purpose 
of the logarithmic amplifier is to compress the preamp output signal with a wide dynamic range to its decibel 
equivalent via a precise nonlinear transformation (Fig.~\ref{fig: amp} right).

\subsection{Transmitter and Receiver Coils}

Our magneto inductance-based system relies on accurately measuring the generated magnetic field 
strength. The initial coil design has high cross-talk voltages between the coils when 
one axe is activated. Depicted in Fig.~\ref{fig: coils} A, the wires bend and do not form a perpendicular structure with regard to the neighboring axes. Due to the lack of distance between the wires of the different 
axes, both inductive and capacitive crosstalk occurs, falsifying and damping the 
output of the active coil (resulting in high discrepancies between the measured and the theoretical magnetic field values). The cross-talks 
on the transmitter side axes result in counter magnetic fields superposing with the aimed generated
magnetic field and therefore changing the overall field layout. Two types of cross-talk decrease the 
performance: Firstly \textbf{inductive cross-talk} being an effect inherent to all systems that utilize alternating 
current. Caused by electromagnetic induction, the level of inductive coupling between two conductors greatly 
depends on their shape, relative orientation and distance. To reduce the effects of inductive cross-talk, 
each coil axis of the new transmitter coil architecture (Fig.~\ref{fig: coils}) consists of two sub-coils whose wiretapping is strictly parallel. The other 
two axes are placed orthogonally to the coil. Secondly, \textbf{capacitive cross-talk} occurs when energy is coupled 
from one circuit to another through an electric field. Regarding two transmitter axes, the excitement of one 
axis (positively charged capacitor plate) also affects the two other coils (negatively charged capacitor 
plates). This influence is strongly reduced by increasing the distance between the axes' wires. Considering these considerations, the new transmitter coil architecture is developed as depicted in Fig.~\ref{fig: coils} D, 
owning a comprehensive advantage in cross-talk, sensitivity, volume, and complexity. Except for improving the inductance value, the new structure has an obvious advantage after the modification. As listed in Table~\ref{tab:table1}, the average cross-talk decreased from more than 60\% to less than 10\%.

\begin{table}
\caption{Cross-talk Voltage in Each Coils}
\label{tab:table1}
\begin{center}
\begin{tabular}{ p{1.8cm} p{0.8cm} p{1.2cm} p{1.2cm} p{1.2cm}}\hline
Transmitter Coil  & Activated Axis & Voltage on X  & Voltage on Y & Voltage on Z \\\hline
 & X & 232 & 155 (63\%) & 137 (58\%) \\
Old / Cube& Y & 158 (68\%) & 245 &179 (67\%)\\
& Z & 132 (63\%)& 167 (68\%)  &236 \\
& \textbf{Avg.} & \textbf{57\%} & \textbf{66\%} & \textbf{63\%} \\\hline
 & X &286 &21 (8\%)  &30 (11\%) \\
New / Spherical & Y & 17 (6\%)  &270 &24 (9\%) \\
& Z &31 (11\%)  &27 (10\%)  & 272 \\
& \textbf{Avg.} & \textbf{9\%} & \textbf{9\%} & \textbf{10\%} \\\hline
\end{tabular}
\end{center}
\end{table}

The same effect is visible on the old receiver side coil setup. To reduce this effect, we restructured the layout and spatial placement of the receiver coils. We evaluated different sizes of receiver coils and their characteristics with regard to covered area and size to support wearability.  
\begin{figure}[hbt]
\centering
\includegraphics[width=0.5\textwidth]{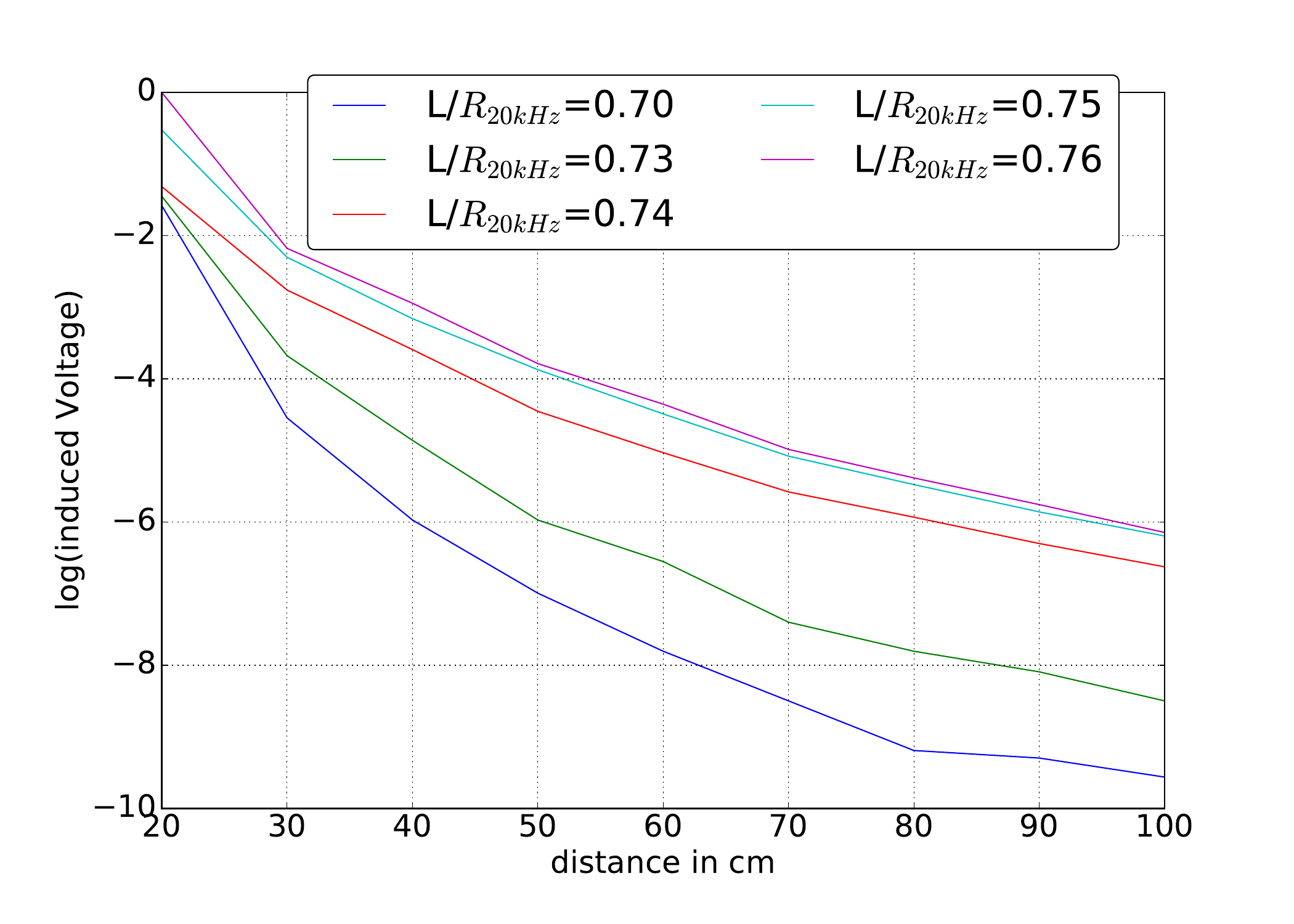}
\caption{Induced voltages of different receiver coils with regard to distance.}
\label{fig: rec range}
\end{figure}
A set of 5 different receiver coil types have been evaluated concerning sensitivity and range. A transmitter coil was constantly excited using a function generator (20kHz, 20 V excitement voltage), and the receiver coil was moved along the transmitter axis in 10 cm steps. We recorded the induced voltages. Results can be seen in Fig.~\ref{fig: rec range}. Considering this measurement's results, we chose the 5th coil with the highest $L/R$ ratio as the receiver coil.

\begin{figure*}[hbt]

\begin{minipage}[t]{0.5\linewidth}
\centering
\includegraphics[width=0.99\textwidth,height=6.5cm]{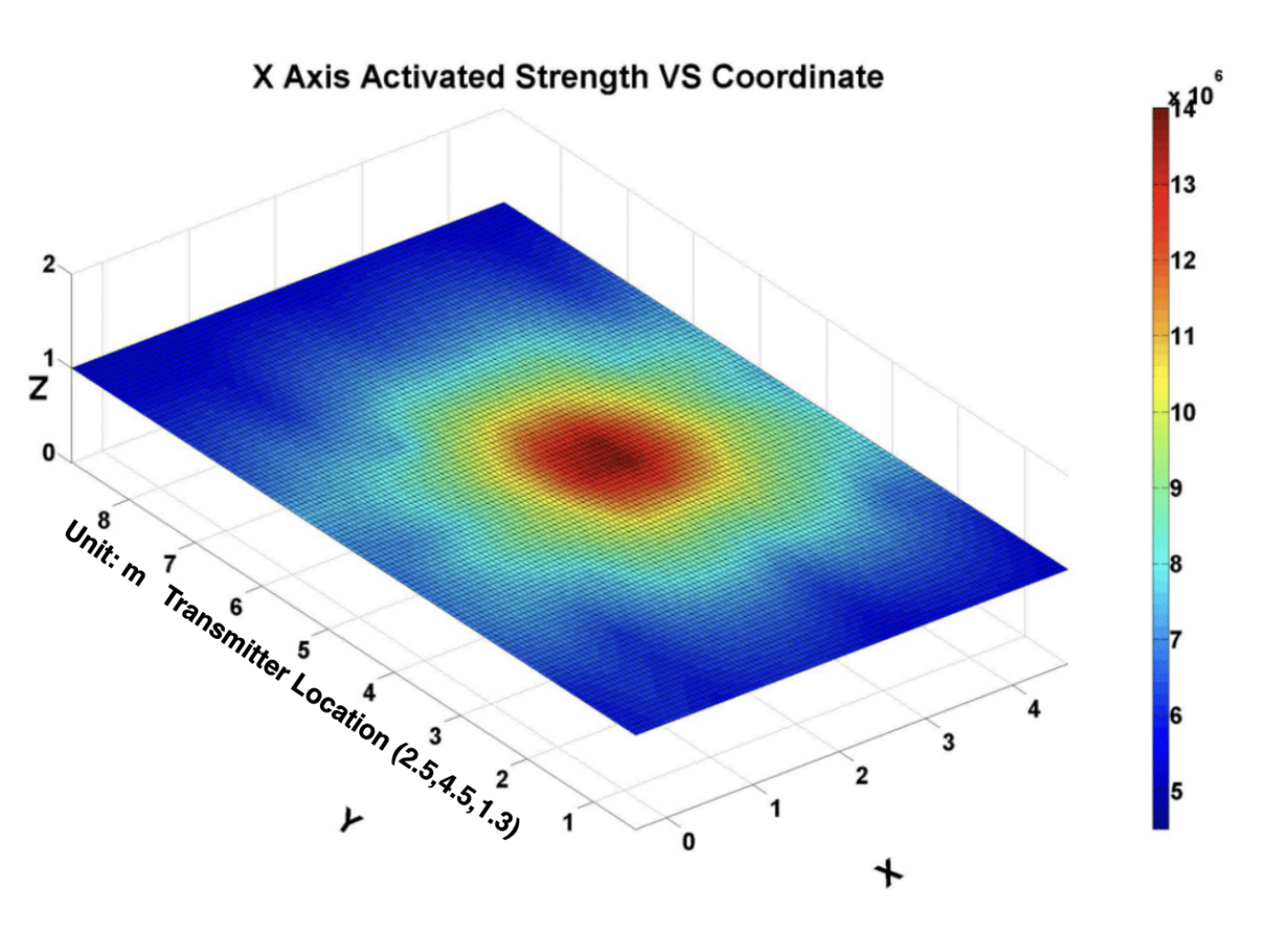}
\caption{X Axis Activated Magnetic Field Strength VS Coordinate(First Experiment)}
\label{fig:figure6}
\end{minipage}
\begin{minipage}[t]{0.5\linewidth}
\centering
\graphicspath{{./Figures/}}
\includegraphics[width=0.99\textwidth,height=6.5cm]{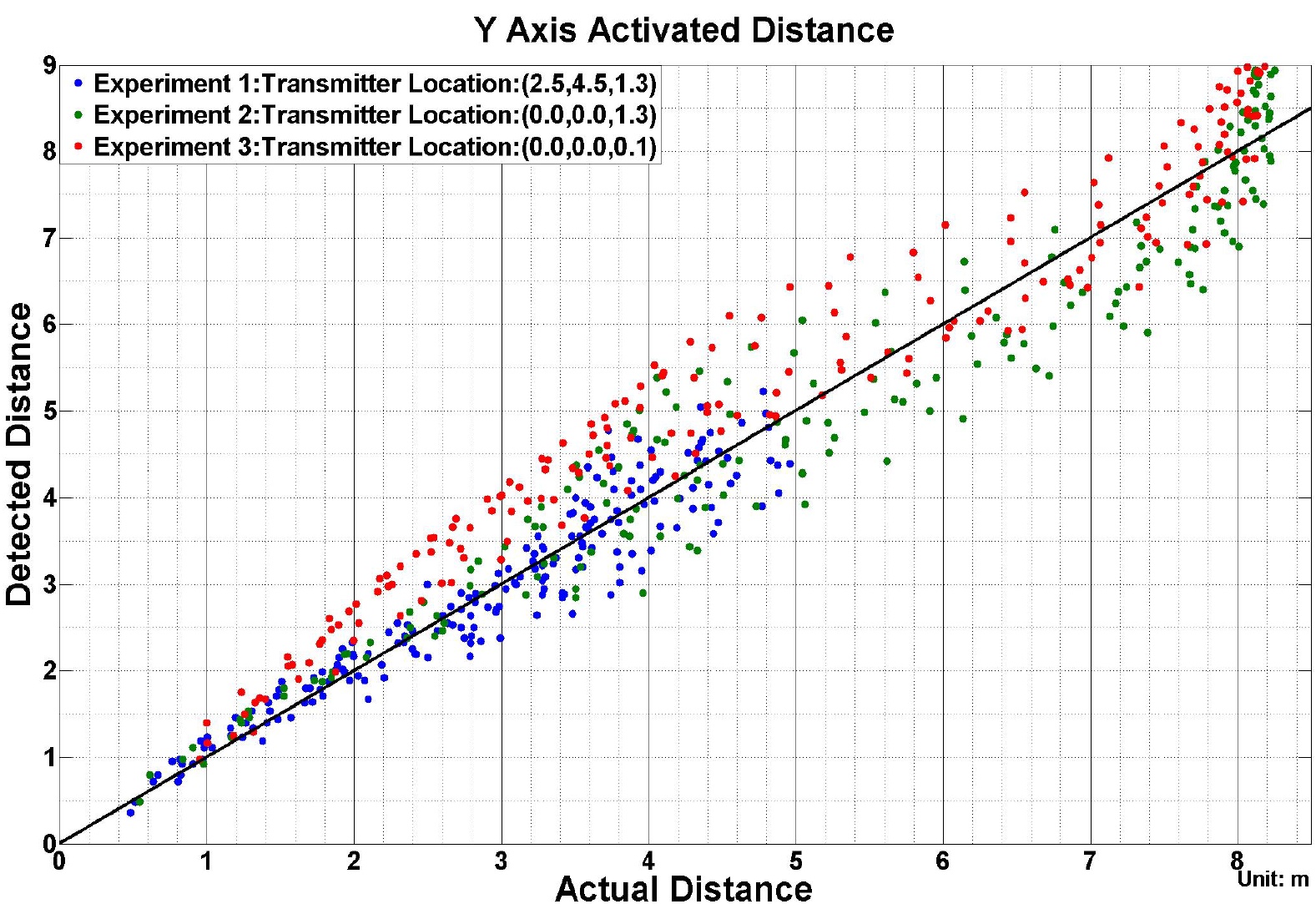}
\caption[width=1cm,height=1cm]{Field Transmission from Transmitter with Y Axis Activated}
\label{fig:figure7}
\end{minipage}
\end{figure*}

\section{Single Transmitter Evaluation}
To verify the performance of the developed system, 4 groups of experiments are 
carried out with a single transmitter within a building. Three of them are conducted to 
evaluate the effects of static  (concrete iron reinforced) building elements on the distance 
accuracy of the system. Thus, the transmitter is placed separately at the center as well as at 
the corner of the testing grid at different heights. The fourth experiment addresses the 
influence of furniture on the magnetic field. A 0.5m grid in an area of 5 $\times$ $8.5m^2$ is 
built, and the field strength values are collected from 197 spatial points to verify the performance 
of this system. 

After data recording, the data is used to determine calibration values ($a_i$ and $b_i$) for each axis of the following transformation function, which is used to transfer the measured magnitude values into distance values:
\begin{equation}
d_i=\sqrt[b_i]{a_i \Vert B_i\Vert}
\end{equation}
The equation is derived from equation~\ref{eq: magnetic} (including simplifications); it combines the transmitter coil variables (number of windings and applied current) in addition to hardware-related variables (damping, resistance of the coil).
After the calibration values are estimated, the transfer function is applied to the measured values to retrieve the distance between the transmitter and the receiver.

\begin{figure*}[hbt]
\graphicspath{{./Figures/}}
\centering
\includegraphics[width=1.0\textwidth,height=6cm]{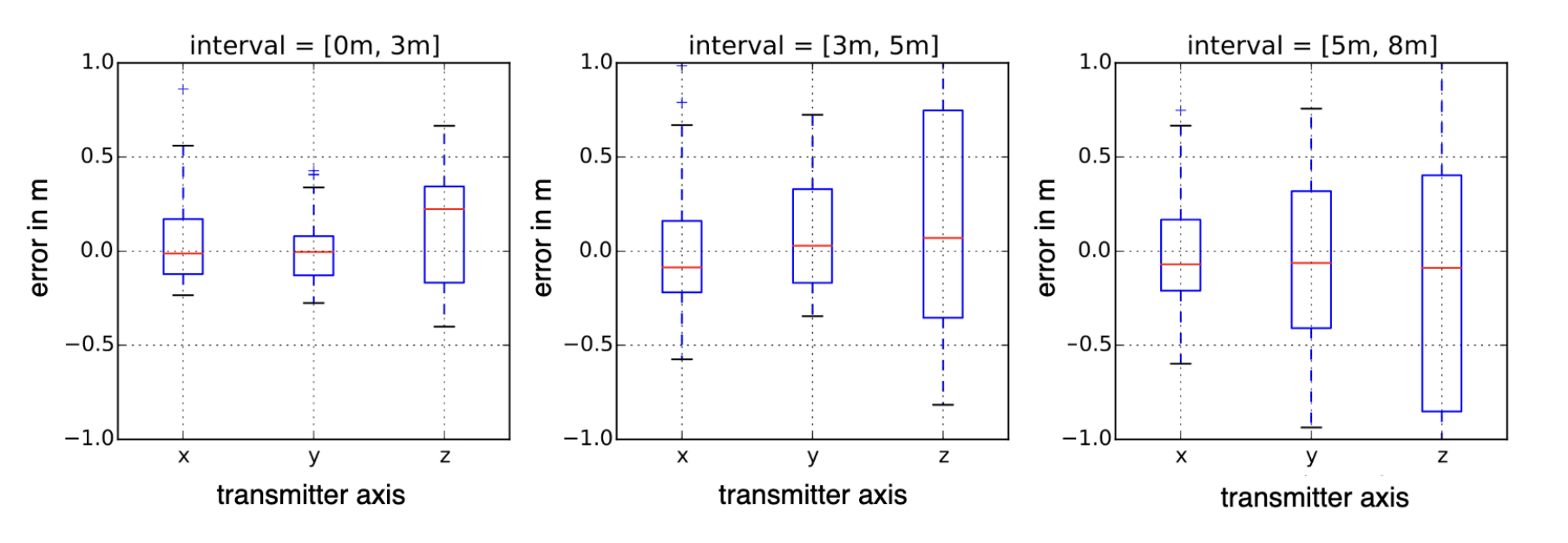}
\caption{Boxplots of Errors for Different Distance Intervals}
\label{fig:figure8}
\end{figure*}

Fig.~\ref{fig:figure6} describes the distribution of magnetic field strength in the room, and Fig.~\ref{fig:figure7} illustrates the actual distance vs detected distance in the first three experiments, in which a single axis of the transmitter is activated. The maximum measuring distance reaches up to 8 meters. Also visible is the increased influence of environmental effects with rising distance as conductive materials generate counter magnetic fields, superposing the generated magnetic field and, therefore, falsifying the measurements.
The discrete level of the collected data is presented by RMSE (root mean square error); the accuracy is listed in Table~\ref{tab:accuracy distance} when Y-Axis is activated. The accuracy is below 0.20m in the distance interval smaller than 3m in the first two experiments and below 0.5m in the third experiment. This also demonstrates that putting the transmitter near the ground hurts the distribution of the activated magnetic field as the concrete floor changes the magnetic field. In the range of 3-5m, The accuracy drops from 0.374m to 0.896m. In the range of 5-8m, the accuracy is 0.9m on average when the transmitter has more surroundings made of concrete.
\begin{table}
\caption{Accuracy with Y-Axis Activated Magnetic Field}

\begin{center}
\begin{tabular}{|c|c|c|c|c|c|}\hline
Experiment & Error in m (0-3m) &   Error in m (3-5m) &  Error in m (5-8m)\\ \hline
 1 & 0.166 & 0.374 & space limitation \\ \hline 
 2 & 0.176 & 0.370 & 0.487\\  \hline 
 3 & 0.445 & 0.896 & 0.750 \\ \hline 
\textbf{Avg.} & \textbf{0.393} & \textbf{0.547} & \textbf{0.619} \\ \hline 
\end{tabular}
\end{center}
\label{tab:accuracy distance}
\end{table}
Fig.~\ref{fig:figure8} illustrates the accuracy of each axis in different intervals 
in the second experiment with the location of the transmitter in the center of the testing 
grid (0.0, 0.0, 1.3). High error rates in the third experiment arise as the 
the transmitter coil is placed on the concrete floor resulting in mutual EM reactions 
and reduced magnetic fields. Main sources of influence of the test environment 
(kitchen appliance, Refrigerator, coffee machine and TV) are in the distance 
interval of 3 to 5m, free areas are between 0 to 3m and 5 to 8m.
Due to the falling signal-to-noise ratio with rising distance, the influence of environmental 
noises rise with increased distance, which is visible in the larger error. 
Another phenomenon is that the 
three magnetic fields induced by three perpendicular coils distribute nearly even in the first 3 meters, 
but from 3 to 5 meters, the three magnetic fields show different accuracies, which are also caused by 
environmental effects, especially walls and floor. From 5 to 8 meters, the strength of the Z field becomes 
weaker, and the accuracy also decreases because the Z field is much closer to the floor and ceiling 
than the other two fields, which have a longer distance to walls, and, therefore, their accuracies are much better. 
As described above, the transmitter coil close to the iron-reinforced concrete floor reduces the system's accuracy as the shape of the magnetic field is altered by the iron reinforcement.

\begin{figure}[t]
\centering
\includegraphics[width=0.50\textwidth]{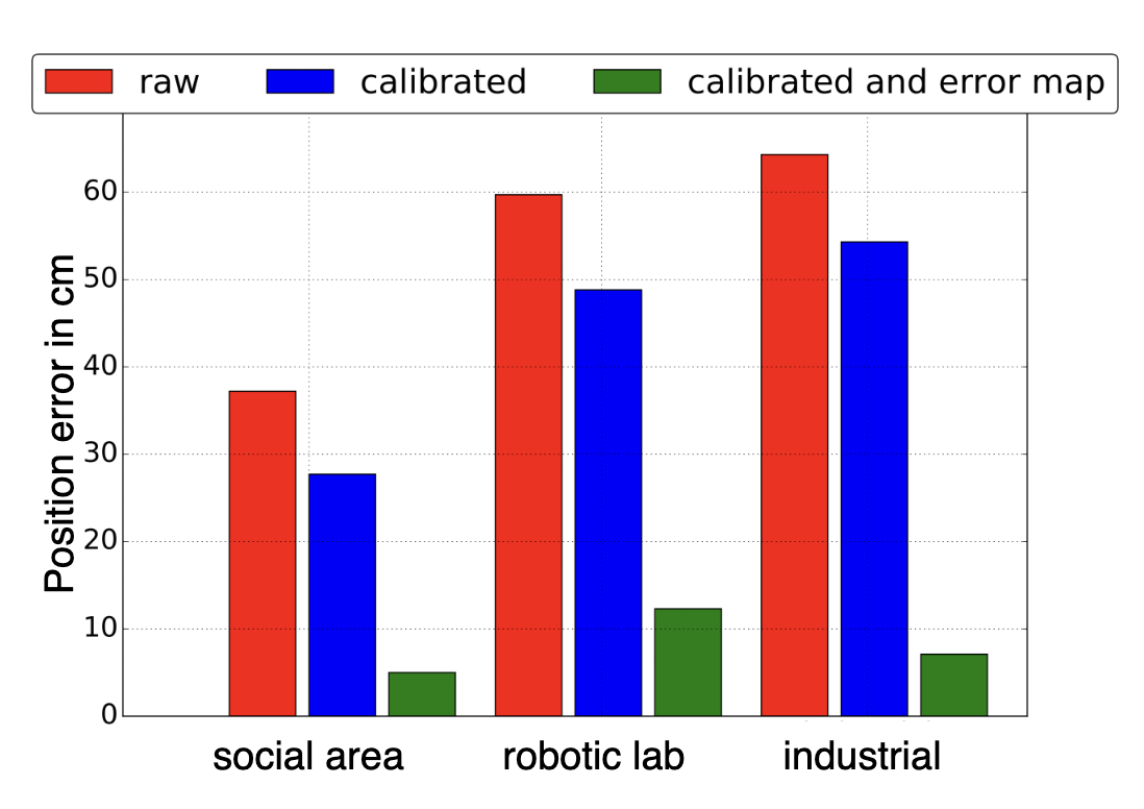}
\caption{Median position errors of the different environments and approaches.}
\label{fig: result}
\end{figure}

\begin{figure*}[t]
\centering
\includegraphics[width=1.0\textwidth]{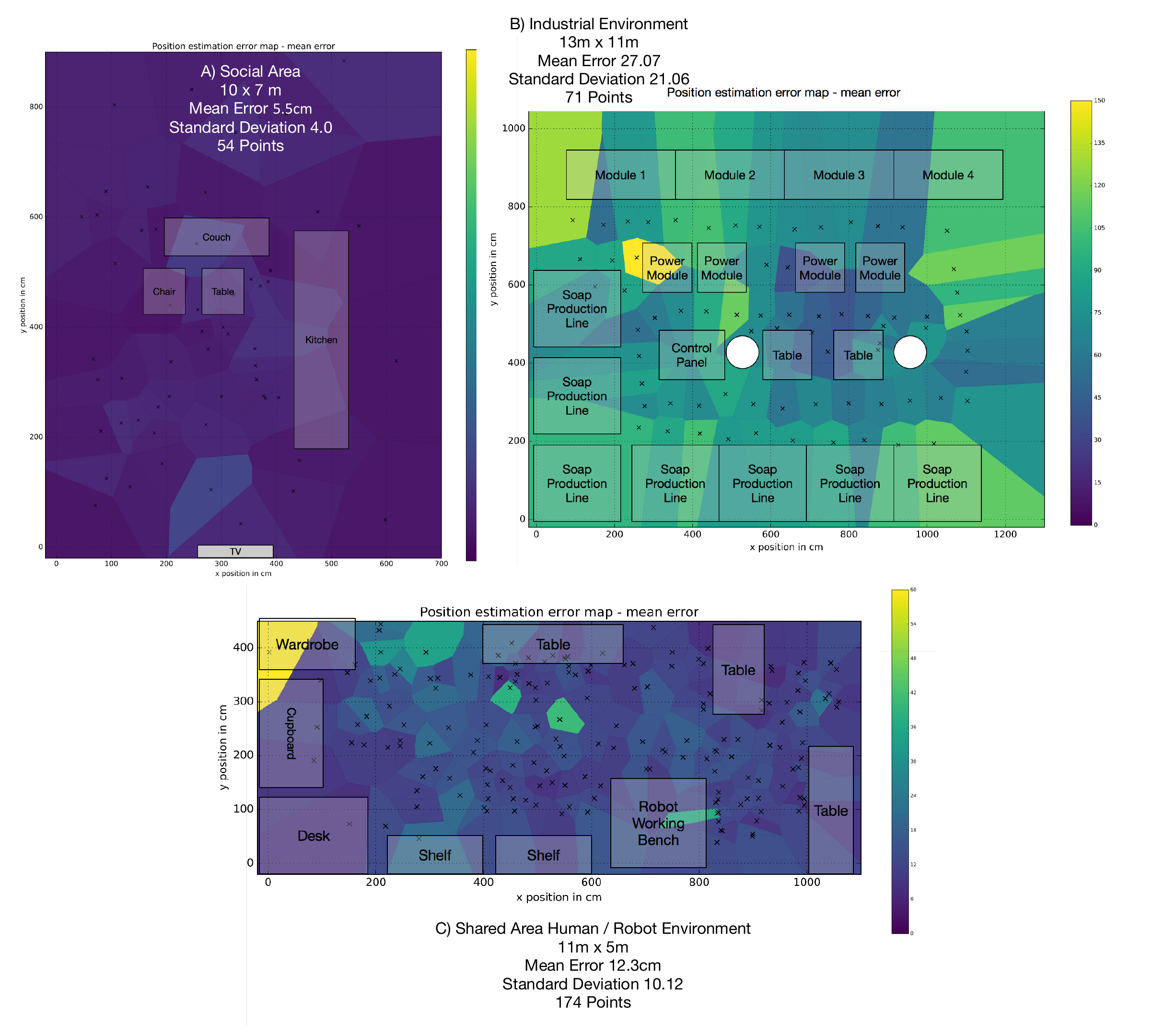}
\caption{Evaluation of the system in social area, industrial environment, and a shared robotic-human environment.}
\label{fig: bremen}
\end{figure*}

\section{Positioning Evaluation}
Our distance estimation approach using equation~\ref{eq: magnetic} supports a calibration process 
relying on only a small number of reference positions to determine the two calibration values $a_i$ 
and $b_i$ for each transmitter axis. The distance error depends on the distance between the transmitter and the receiver (see table~\ref{tab:accuracy distance}).  
Building upon Pirkl's work, we combine measurements between multiple stationary transmitters 
and the mobile receiver. Starting from this approach, we conducted data recordings in three 
environments, an office area with couch and kitchen elements ($10m\times 7m$), a laboratory with 
stationary robot arms and mobile robotic rovers ($13m\times 11m$), and an industrial production line 
($13m\times 11m$) which holds high-power stepper motors for the conveyor belts and metal production 
modules. The magnetic field localization system uses seven transmitters in all these environments, and the anchors 
are placed at the corners of the environments and - if possible - in the center of the areas to provide enough 
anchor information for a distinct position estimation. We recorded all in all around 300 different positions in these areas. The raw magnetic field measurements are transformed into distances as described by equation~\ref{eq: magnetic} with calibration values from the initial data recordings.
A standard triangulation algorithm (intersection of circles) then calculates the position of the receiver 
relying on the distances to the transmitters.  

The results of the data recording and the applied algorithms are shown in Fig.~\ref{fig: result}.
In hazardous environments as for example in industrial production lines or rooms with 
human/robotics interaction, metal objects or electro-magnetic 
sources influence the measured 
magnetic fields. It also locally influences the 
transmitters' calibration values $a_i$ and $b_i$. To overcome this, we use 10 
percent of each recorded data set to recalculate the calibration values to the 
local environment. This results in new calibration values $a_i^{local}$ and $b_i^{local}$. 
Applying these calibration values to the distance and position estimation algorithms 
results in lower - but still unsatisfactory position estimation results. The main 
reason for these errors arises due to local influences of metal objects and EM sources. 
The local layout of the magnetic field is jolted, altering the fields in either higher or lower 
magnetic field strengths and, therefore, corresponding falsified induced voltage readings. Although 
creating a discrepancy between measurement 
and the theoretical magnetic field model, the discrepancy is repeatable for stationary 
objects. Relying on the repeatable distortions caused by obstacles, we algorithmically 
compensate for this influence by creating a lookup map comparable to the fingerprinting 
approach. We randomly choose 30 percent of the data sets to generate the error map. The 
fingerprinting map uses the magnetic field vectors of a measurement cycle (all 
transmitters have generated their magnetic fields) to describe the receiver's position. A regression algorithm approximates 
the position values and allows non-discrete position lookups. 
The result of this is presented in figure~\ref{fig: result}. For data sets with 
lower number of positions, the LASSO Algorithm results in better performance, for 
a higher number of estimation points as e.g. in the industrial or robotic environment, 
the influences of the obstacles are approximated in a better way by polynomial 
regressors with degree 3, taking the nature of the magnetic field degradation against 
distance into account. Fig.~\ref{fig: bremen} depicts the Voronoi position error diagrams and the reference position distribution of the different environments. The industrial environment has the highest position error as many conductive materials disturb the measurements, resulting in higher position errors.

\section{Conclusion}
This paper presents hardware-related optimizations made on the initial circuits presented by Pirkl et al.  Focusing on a more specialized coil design on the receiver and the transmitter side, the influence of inductive and capacitive cross-talk has strongly been reduced. In addition, the sensitivity of the receiver-side analog circuit has been increased. Those improvements pushed the maximum range of the positioning system from 4m to 8m ($50m^2$ to $200m^2$). The standard deviation of the distance errors drops from initially 0.56m to 0.20m in the near field. Using a fingerprinting approach to overcome and deal with local magnetic field distortions reduces positioning error to below 30 cm in industrial environments.

\bibliographystyle{IEEEtran}
\bibliography{reference}{}

\end{document}